# An Early Warning System for Bankruptcy Prediction: lessons from the Venezuelan Bank Crisis

Financial Markets. Investments and Risk Management

**Loren Trigo**

Professor

Center for Production and Technological Innovation
Instituto de Estudios Superiores de Administración
Calles IESA, San Bernardino, Caracas, Venezuela
58 212 555 4498
58 212 555 4494 fax
loren.trigo@iesa.edu.ve

**Sabatino Costanzo**

Professor

Center of Finance
Instituto de Estudios Superiores de Administración
Calles IESA, San Bernardino, Caracas, Venezuela
58 212 555 4498
58 212 555 4494 fax
sabatino.costanzo@iesa.edu.ve

# An Early Warning System for Bankruptcy Prediction: lessons from the Venezuelan Bank Crisis

*Abstract*

During the years 1993-94 Venezuela experienced a severe banking crisis which ended up with a total of 18 commercial banks intervened by the government.  Here we develop an early warning model for detecting credit related bankruptcy through discriminant functions developed on financial and macroeconomic data predating the 1993-94 crisis.  A robustness test with data from 1996-2004 performed on these functions, shows high precision in Type 1 & 2 errors. The model calibrated on pre-crisis data is even able to detect abnormal financial tension in the late Banco Capital by June 2000, many months before it was intervened and liquidated.

*Introduction*

**About banking crises**

There are many definitions of what constitutes a banking crisis.  Sundararajan and Baliño (1991) mention some of the definitions in the relevant literature.  Some examples of these definitions are the following:

- "A money demand that cannot be satisfied by all participants in the short run" Schwartz (1985); Mirón (1986); Wolfon (1986)

- "Liquidation of loans that were granted during an economic boom" Veblen (1904); Mitchel (1941)

- "Circumstance in which the suppliers of lending funds that in usual conditions are able to grant loans without any difficulties find themselves unable to grant these loans in any form or term" Guttentag and Herring (1983), Manikow(1986)

In addition, Morón and Aguero (2003) refer to the following concept in their work:

- "A banking crisis is a circumstance where a bank or a group of banks becomes insolvent, with the consequence that the value of their assets falls below the value of their obligations" Bell and pain (2000)

**Main microeconomic causes of banking crises**

*Disproportionate increase of the size of the credit portfolio*

Most banking crises follow an economic boom (García 1998), often accompanied by an increase in the size of the credit portfolios of banks.  In most of these cases, according to Barton, Newell and Wilson (2002), these new credits are often bad credits that can ignite a crisis.

*High number of expired loans*

Given a booming credit portfolio plagued with problematic loans, the portfolio of expired loans increases in two ways: there is an increase in the number of overdue loans and of loans under litigation.  This of course affects the balance sheet; nevertheless, very often this situation goes unnoticed because the banks dress up their financial statements to hide their expired loan portfolio problems.

*High number of highly correlated loans*

Two loans are correlated if their collaterals are correlated.  With respect to stock market bubbles, according to Barton, Newell and Wilson (2002), several years of continued growth in the price of stocks can indicate the existence of a stock market bubble that could burst.  This is dangerous because 1. Many loans are backed up by shares of stock and a stock market crash will cause a reduction in the collateral of these loans and 2. Many banks invest in stocks as well and stand to lose that investment if markets become deeply bearish.

## **Main Macroeconomic Causes of Banking Crises**

*Economic instability*

According to García (2000), one of the main causes of banking crises is an economic growth that is relatively unstable or with very pronounced cycles covering short time periods. These cycles are not infrequent in oil exporting countries with oil price fluctuations that make their economies highly volatile. One of the problems derived from these cycles is the abnormal growth of lending portfolios during intervals of economic expansion, with good loans being granted at the beginning of the good times and bad loans thereafter, resulting an a deteriorated lending portfolio whenever the good times end up being shorter than expected. The crucial point of this argument is that this deterioration is not visible except after the economic slowdown, when the losses within lending portfolio begin to materialize.

Previous economic deregulation

When interest rates are regulated it is not possible for banks to increase their client base by attractive lending rates that differ significantly from the regulated rate. Demirguc-Kunt and Detragiache(1998) report that banking crises are more probable in deregulated financial systems. They document a direct relation between the number of banking bankruptcies and the weakness of regulatory control and the inefficacy or negligence of supervisory entities. Kaminski and Reinhart (1995) also point out the coincidence of banking and stock market crises in deregulated economies.

*Occurrence of a boom in the years prior to the crisis*

According to Kindelberger (1978), "the Financial Crises are associated with the peaks of certain economic cycles… A financial crisis is the culmination of a period of economic expansion that is about to end". During the boom, the crisis is prepared by the uncontrolled growth of the lending portfolios. Then the very growth of the lending portfolios cause a hike in lending interest rates to the point where the demand for credit fades away. The higher interest rates can certainly cause a contraction of the economy. A plunge in the price of bank assets and credit collaterals that safeguard the lending porfolios can in turn expose lending institutions to bankruptcy risk.

*Fall in the demand for money*

Negative shocks and expectations of future devaluations produce reductions in the demand for money because economic agents try to exchange their positions in local currency with equivalent positions in foreign currency or assets. These events occurred in Venezuela in the years preceding the banking crisis, due to the political instability caused by two attempted coups d'etat in 1992 and to the deep depression caused by the government's budgetary deficit, the fall of the local stock and real estate markets. This fall in the demand for money is very prejudicial for banks because it results in a steep reduction of the banking sector's primary resource: the size of the accumulated money mass deposited by bank customers. At the same time, because of the decrease in the demand for money banks cannot replace quickly enough old expired loans with fresh loans that would warrant the continued profitability of the loan portfolio.

*Chronic government budget deficits*

Chronic governement budget deficits are accompanied by interest rate hikes and inflation. "Due to the use by certain countries of the exchenge rate as nominal anchor or inflation fighting tool, the presence of chronic budget deficits contributed to the real appreciation that led to problems in the balance of payments and later to banking crises."

## **Other Causes of Banking Crises**

*Contagion Effects and Bank Runs*

According to Sundarajan and Baliño (1991), the contagion effect in the banking sector occurs when following the bankruptcy of a given bank, the growing anxiety of the savings account owners affects other banks of reputation similar or worse than that of the bank that went bankrupt. This phenomenon manifests itself in the form of bank runs by savings account owners on banks suspected of being in financial trouble, bank runs that could result in the latter's bankruptcy.

*Faults in Bank Regulation and Supervision*

To avoid the effects of a systemic or isolated banking crisis it is important that the entity in charge of supervising the banking industry carry out its function in an effective manner. The relevant literature indicates that in many banking crises were made worse by a complacent, irresponsible supervising entities. The manner or way in which the supervising entity reacts to a banking crisis is also important: indecisiveness may highten public distrust or fear and an ever worsening contagion effect.. During the 3-month long "closed-doors" intervention of Banco Latino in 1993-94, the supervising entities (SIB, Fogade, Ministerio de Hacienda and BCV) gave the impression of being unprepared to deal with a banking crisis of that magnitude. This impression grew into a panic that infected most of the banking system and resulted in the bankruptcy of 18 banks.

**Consequences of the Banking Crisis**

- The consequences of the banking crises are numerous. Morón and Aguero (2003) cite the following:

- They cause a great cost to the state in the recovery of deposits of the public (in countries with insured deposits) because of the huge tranfer of funds to the bankrupt banks that must meet their obligations with their savings account owners.

- They affect the system of payments, which is one of the main functions of banking institutions. As a result, transactions become costly for all economic agents.

- They cease to function as financial intermediaries, therefore many productive proyects fail to be realized for lack of resources.

- In case there is no deposit insurance, a great social cost results from the loss of savings. These social costs trigger political tensions which if mishandled can degenerate into grave chaos.

- The economy stops growing. Relevant literature shows that economic growth after a banking crisis is often negative.

- Banking crises reduce the effect of monetary policies.

*Can a banking crisis be predicted?*

According to Berg and Pattiño (2000), the integration of the world markets, the velocity with which money moves from one country to another, the globalization and the opening of capital make more complex the task of predicting banking crises. In their study entitled *"Dificultades para la predicción de crisis económicas",* Berg and Patiño argue that the banking crises in Mexico, Korea, Thailand, Indonesia and Malasia share common causes, despite their exhibiting great differences that make difficult the construction of a unique model of early crisis alert.

For example, in Thailand the main variables were

- the ratio of deposit account deficits to the GNP
- the variation in foreign reserves

whereas in Malasia the main variables were

- the ratio of the deposit account deficits to the GNP
- the variation in international reserves
- the over-valuation of the exchange rate.

Finally Berg and Patiño study the crises in Korea and Indonesia and find a key role for
 The ratio of the short term foreign debt and foreign reserves

Likewise, Kaminsky (1998) considers that in Latin America and in Thailand a key role is played by the great expansion of credit in the banking system during the years prior to the crisis. He also argues that in Chile the banking crisis was to a great extent oriented to the foreign sector. In Colombia a key role was played by

- the m2 multiplier,
- the domestic credit index with the GNP and
- the m2 index on foreign reserves.

Berg and Pattiño (2000) conclude their work arguing that the models of prediction for banking crises must never be used in isolation, but must be used with other suitable tools.

## *Brief summary of the 1993-1994 banking crisis in Venezuela*

From 1992 on the fragility of the banking system in Venezuela begins to show. At that time Venezuela had just come out of an economic Boom produced by the increase in the price of oil following the onset of the Gulf War (1990-1991). Then in February and november of 1992 two attempts of coup d'etat occurred in Venezuela, both of which failed, creating a climate of extreme economic, political and social instability. According to García (1998), prior to 1992 the Venezuelan Central Bank (BCV) carried out a number of studies of the banking industry and had alerted the public regarding the fragility of that sector. In 1993 the BCV informed the executive branch of the insolvency of a number of banking institutions. Despite these warnings, the oversight agencies did not act quickly and thus allowed some banks to continue running deficits.

On December 22, 1993, the president of Banco Latino, the second largest bank in the country, resigns following rumors about the precarious situation of the bank's position. On January 12 of 1994 Banco Latino leaves the Bureau of Compensation and the next day is intervened by the government behind closed doors. As was pointed out above, the magnitude of a banking crisis depends in large measure on the way in which supervisory and regulatory agencies react. Given this, the intervention of Banco Latino behind closed doors created a climate of uncertainty and panic among thousand of account owners of this financial institution. When the public began to suspect that there were other banks in similar or worse conditions, a series of bank runs occurred, and the contagion effect became generalized.

The regulatory and supervisory agencies' lack of capacity to deal with the impending crisis became evident. During the first months of the crisis a series of contraditcions plagued the agencies that were responsible for attending to the crisis; namely, the Venezuelan Central Bank, Superintendencia de Bancos, Consejo Bancario Nacional, Fogade and the Ministry of Finance.

In june of 1994, the supervisory agency could no longer able to keep a secret that it was intervening Banco Maracaibo, Construccion, Metropolitano, La Guaira, Amazonas and Bancor.

In these interventions they committed more mistakes, among which Garcia(1998) point out:

- The granting of financial help under the assumption that the basic problem was one of lack of liquidity (and not one of insolvency).

- The granting of help without removal of management. In many cases these cash injections wound up in related accounts, offshore banking accounts and foreign currency.

Some time after, these institutions were closed and liquidated anyway.

In August of 1994 new interventions occurred, this time affecting Banco Venezuela, Consolidado, Progreso, Italo-Venezuelan, Profesional and Principal. But when Banco de Venezuela and Consolidado were intervened, a new open door model was put into practice: the institutions were disinfected, bank management was removed from office and the banks were sold to interested investors. A similar model was used with Banco Progreso, but when the confidence of account holders was not recovered –among other things, the president of Banco Progresso was out of the country—the agency decided to tranfer the accounts to state owned institutions. The rest of the banks intervened that month had the same fate as Banco Progreso: they were liquidated and closed.

At the end of 1994 and beginning 1995, Banco Andino, Empresarial, Union and Principal were intervened, from which the first and third were liquidated and closed, while the second was forced to look for a partner that would give it capital or else sell off part of itself. Despite this, the bank continued having solvency problems and was hard pressed by account owners and by competent agencies to the point that the president had to travel to Colombia to a meeting with the directives of Banco Ganadero in order to obtain written statement of intent to buy. The merger never took place but the news caused the situation to stabilize and the bank runs ceased.

## *Behavior of some macroeconomic variables during the period of the Crisis.*

### *The local interest rate spread*

Although there are various ways of calculating that value, we decided to do so as the difference between the active average rate and the passive average weighted annual rate.
The behavior of this variable should correspond (in theory at least) to the fact that as the crisis approached, risk loving banks would want to rise their passive rate (diminishing the spread) in order to capture funds with which to pay the high levels of debt that they had contracted. Let us see how events occurred in Venezuela during the years 89-93:

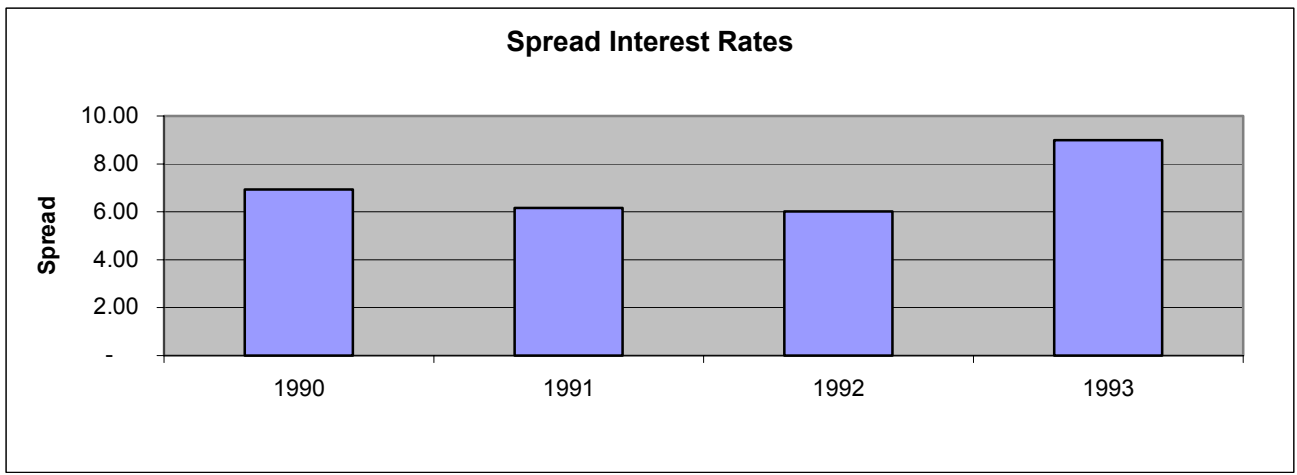

As may be observed, the spread diminishes until 1992 and then bounces back up 50% by the end of 1993. One possible explanation is that the troubled banks –in their need to capture more deposits—rised their passive (outbound) rates and then in order to honor those passive rates had to rise their active (inbound) rates to a level where they sacrificed the quality of their credit portfolio; that is, they allowed their credit portfolio to deteriorate in order to obtain more profits that would allow them to honor their debts.

*The real effective exchange rate*

The money in circulation (m1) is the quantity composed by "coin and bills" and the "deposits in view," whereas monetary liquidity (m2) is equal to the sum of "m1," the "savings deposits" and the "saving certificates." This indicator shows us how in the years previous to the crisis, because of the high passive (outbound) rates, the number of deposits placed in banks increased (specially savings deposits and savings certificates), as one can see in the following graph:

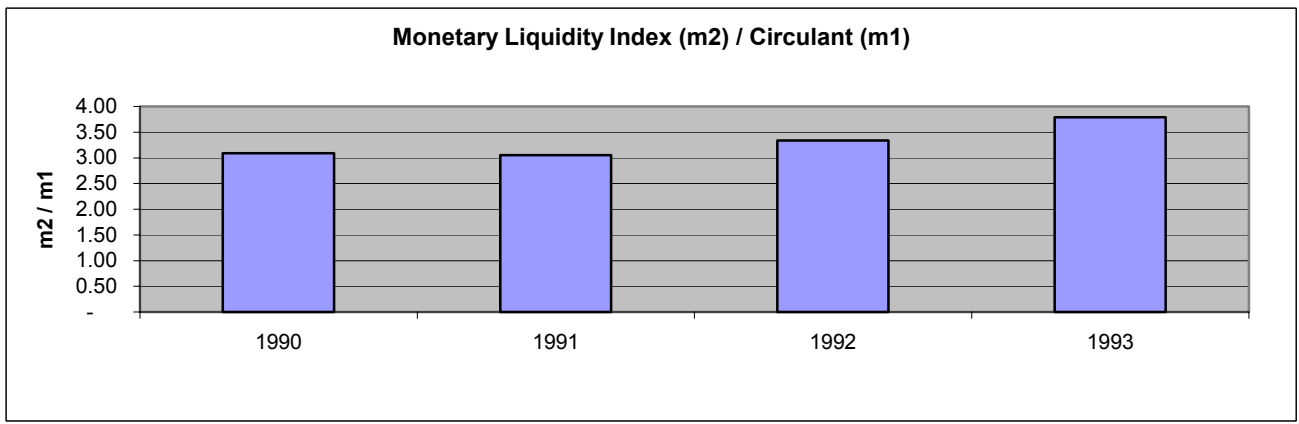

The graph shows that after 1992 this indicator moves up, with a clear preference on the part of the public to maintain savings in banks instead of holding cash.

*Percent change of IGAEM*

As this is an indicator of economic activity, it behaves as the Internal Gross Product in times of banking crises, with a tendency to decrease in the years leading to the crisis, which damages in a significant way the balances of the banks at the time when the economic bubble bursts. Let us see how the IGAEM behaved during the period 1990-1993:

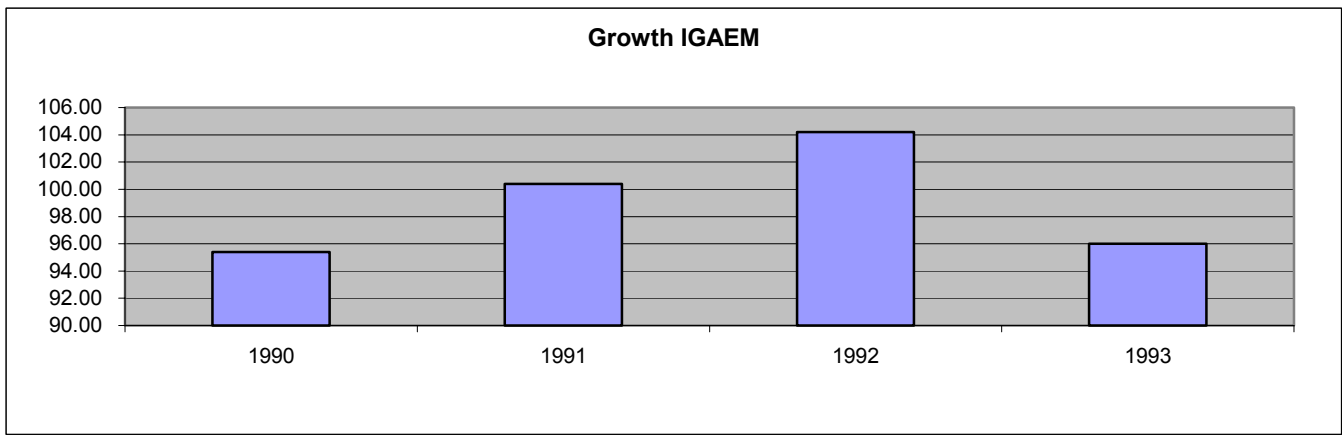

This indicator move just as expected. After a substantial increase in the period 1990-1991 (probably caused by the Persian Gulf War), it began to come down at the end of 1992 (possibly because of the political instability following two attempted coup d'etats). During the year prevous to the crisis, the percent change in this variable was negative.

*Percent change in foreign reserves*

In the years predating the crisis one can see how the level of foreign reserves (minus gold) is diminishing considerably. This drop is due –as was explained in the case of the effective real interest rate—to the flight of capital caused by the social and political upheaval during 1992 and to the perception of a deteriorating balance sheet in the banking sector.

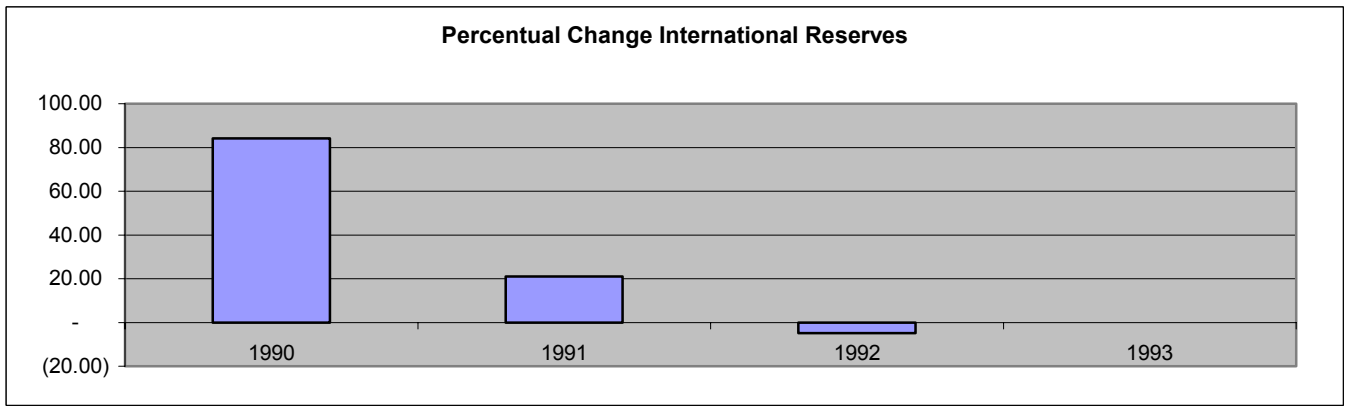

*Methodology*

1. Our model is based on the one Developer by Sanz (1997).
2. Sanz (1997) builds 5 models with discriminant análisis using the financial data from the Venezuelan Banking Crisis previous to 1994. She uses the following variables:

Financial info. on 17 broken banks and 17 not broken 1 year before the crisis.
- Variables: A, B, C, D, E

Financial info. on 17 broken banks and 17 not broken 2 years before the crisis.
- Variables Financieras: F, G, H, I

Financial info. on 18 broken banks and 18 not broken 3 years before the crisis.
- Variables: J

Financial info. on 17 broken banks and 17 not broken 4 years before the crisis.
- Variables: K, L, M

Financial info. on 16 broken banks and 16 not broken, 5 years before the crisis.
- Variables: N, C

**Advantages and limitations of Sanz work (1997)**

*Advantages*

- If we apply the five Sanz models (comparable to 5 different "syndromes") to a given bank, the model according to which this bank falls into bankruptcy (the "syndrome" best fitting the Bank) will also indicate, by construction, the life left to this given bank.

*Limitations*

- These five models are difficult to generalize and even to compare because they are based on different indicators (i.e.: each "syndrome" is a set of "simptoms" that is disjoint with respect to the other "syndromes"). Only the indicator C appears in more than one.

**Indicators used**

- Our model, based in Discriminant Análisis, uses 5 of the financial variables used by Sanz, plus 5 more relevant variables of macro-economic type.
- The financial variables were selected considering the statistical significance found by Sanz(1997) for each one of them.
- All the variables are conceptually justifiable and compatible with the relevant literatura on banks crises.

**Advantages of the new model**

**The incorporation of macro-economic variables allows:**

- The possibility of using a single model along all the different years of the experiment, because these variables capture the time-changes of the context common to all banks. *(notice that for a given bank is much less indicative of structural problems to be in critical condition during a global crisis than during bountiful times).*
- The possibility of evaluating the precision of the model year after year in comparable terms.
- The possibility of extending this model to an international one where we could study banking and money fenomena common to several interconnected but distinctly different economies, as it is the case of a banking crisis contagion among countries whose economies are interconnected.

**Classic financial indicators used in the discriminants model**

- Other assets/ total assets: measures the assets quality.
- Financial outflows/average total inflows: effective passive interest rate, measures the average cost incurred by financial institutions to pay for the resources provided by the public.
- Operative margin/average assets: measures the assets profitability. Reflects the capacity of the institution to carry out normal activities.
- Operative margin/average equity: indicates the equity profitability and the ability of the institution to face its normal operative activities.
- Financial inflows/financial outflows: measures the spread generated by the difference between the rates paid to the public and the ones collected from the credit portfolio and other investments.

M*acroeconomic indicators used in the discriminants*

- Local interest rates spread: the differential between the weighted annual active interest rates and the passive ones..
- Effective real Exchange rate: derived from the inflation adjusted nominal exchange rate.
- Monetary liquidity index (m2) /circulating money (m1): indicates the public preference for savings over cash.
- IGAEM percentual change: percentual change of the Venezuelan Leading Economic Indicator
- Internacional reserves percentual changes (minus gold stock).

*The procedure*

- We calibrate the model on "pre-crisis" data (c. 70 banks at the beginning, reduced to 36 because of data availability and quality reasons)
- We test the model on "post-crisis" data (68 Banks)

**About the "pre-crisis", 1990-1993 data used to <u>calibrate</u> the discriminants model**

- Data from 36 banks:
- Each intervened bank was compared with a non-intervened with similar average capitalization, to obtain 18 intervened banks and 18 non-interveened banks.

**What does it really mean to calibrate the model? A word about the model resulting from the "pre-crisis" calibration:**

After changing the axes of the vector space where each point represents a bank, we obtain the equation of a hiper-plane separating two "Types of banks":

$-0.32 F_1 + 0.20 F_2 - 0.18 F_3 - 0.32 F_4 - 0.99 F_5 + 0.38 M_1 - 0.95 M_2 - 0.5 M_3 - 0.26 M_4 + 0.01 M_5 \geq -190.395$

Where F is a Financial indicator, M a Macro-economic Indicator and -190.395 the bankruptcy "treshold"

In practice, given a bank characterized wby the indicators F1, ... , M5, whenever the weighted sum of its indicators is greater than -190.395, the diagnosis is one of imminent bakruptcy, and vice-versa.

**Linear Discriminant Análisis (Plate)**

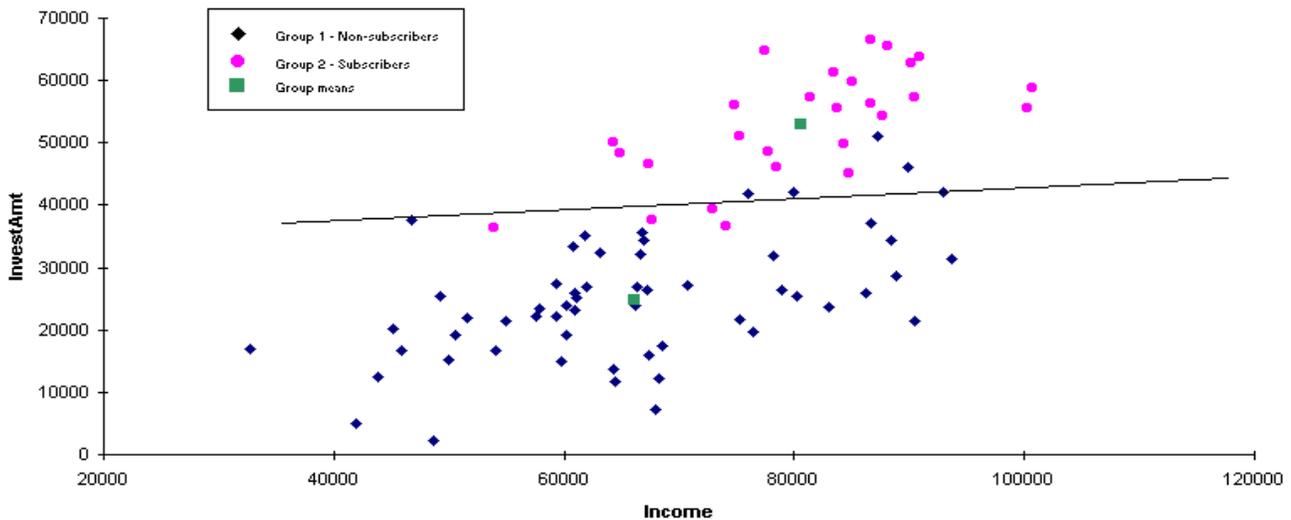

**How the set of "post-crisis", 1996-2004 data is used to <u>test</u> the discriminant model:**

- The formula Developed above is used to "probe" each of the 68 banks every 6 months.

**At last, the indicator gives us a warning about the Banco Capital :**

| Year | Capital | Median | Mean | StDev |
|------|---------|--------|------|-------|
| 2000 | -150 | -294 | -335 | 144 |
| 1999 | -272 | -404 | -464 | 620 |
| 1998 | -270 | -347 | -367 | 111 |
| 1997 | -289 | -445 | -465 | -445 |
| 1996 | -199 | -327 | -346 | 137 |

Banco Capital will be the only one among 68 others breaking the -190.395 threshold in 1996 and then again in June 2000. For the other years, it falls quite close to the threshold. **The Govt. closes Banco Capital on December 2000.**

**The change in Model Precision along the "pre-crisis" years: Adicional Conditions: error of Type I v.s. error of Type II**

*Model obtained from data collected starting 4, 3, 2 and 1 years before the crisis (140 observations).-*

|  |  | 4 years before c. | | | 3 years before c. | | | 2 years before c. | | | 1 year before c. | | |
|---|---|---|---|---|---|---|---|---|---|---|---|---|---|
|  |  | Yes | No | % | Yes | No | % | Yes | No | % | Yes | No | % |
|  | Yes | 58 | 12 | 83 | 47 | 6 | 89 | 34 | 1 | 97 | 16 | 1 | 94 |
|  | No | 22 | 48 | 69 | 18 | 35 | 66 | 12 | 23 | 66 | 7 | 10 | 59 |
| **Accuracy** | | **75.71%** | | | **77.36%** | | | **81.43%** | | | **76.47%** | | |

*Note:*
- If we require that Error of Type I (bankruptcy omission) to be less than the Error of Type II (false bankruptcy alarm), the model generates false alarms in a 10% of the "post-crisis" data.
- This is compatible with the relevant literature: it is well known that most of the Early Warning Systems often generate "false alarms".

## Bibliography

- Berg, Andrew y Pattiño, Catherine, *Dificultades para la predicción de crisis económicas,* Fondo Monetario Internacional Washington, 2000.

- Dominic Barton, Roberto Newell y Gregory Wilson, *Dangerous Markets: Managing in Financial Crises.* 2003.

- García Gustavo, *Lecciones de la Crisis Bancaria de Venezuela,* Ediciones IESA, 1998.

- Kaminsky, Graciela y Reinhart, Carmen, *The Twin Crises: The Causes of Banking and Balance-of-Payments Problems,* American Economic Review, 1995.

- Kaminsky, Graciela, *Currency and Banking Crises: The Early Warnings of Distress,* Board of Governors of the Federal Reserve System, 1998.

- Kindelberger Charles, *Manias, panics and crashes,* New York, NY: Basic Books Inc, 1978.

- Minsky, Hyman P., *The Financial-Instability Hypothesis: Capitalist Process and the Behavior of the Economy,* Cambridge University Press, 1982.

- Miron J.A., *Financial Panics, the Seasonality of the Nominal Interest Rate, and the Founding of the FED,* American Economic Review, 1986.

- Morón Eduardo y Aguero, Loo-kung, Sistema de Alerta Temprana de Fragilidad Financiera, Universidad del Pacifico, 2003.

- Paracare Elsy y Zany Victor Fermín, Un Indicador Mensual de Actividad Económica (IGAEM), Serie Enero 2000 del Banco Central de Venezuela.

- Schwartz, Anna, Real and Pseudo financial crises, New York: San Martin's Press, 1985.